\documentclass[journal]{IEEEtran}
\usepackage{graphicx}
\usepackage{epstopdf}
\usepackage{float}
\usepackage[table]{xcolor}
\usepackage{ulem}
\usepackage{soul}
\usepackage{amsmath}

\usepackage{bm}

\ifCLASSINFOpdf
\else
\fi
\begin{document}
%
%
%
%

\title{A Physics Based Multiscale Compact Model of  \textit{p-i-n} Avalanche Photodiodes\\
}
\author{\IEEEauthorblockN{Sheikh Z. Ahmed\IEEEauthorrefmark{1}, Samiran Ganguly\IEEEauthorrefmark{1},
Yuan Yuan\IEEEauthorrefmark{2},
Jiyuan Zheng\IEEEauthorrefmark{3},Yaohua Tan\IEEEauthorrefmark{4}, Joe C. Campbell\IEEEauthorrefmark{1} and Avik W. Ghosh\IEEEauthorrefmark{1} \IEEEauthorrefmark{5}}\\
\IEEEauthorblockA{\IEEEauthorrefmark{1}Dept. of Electrical and Computer Engineering, University of Virginia, Charlottesville, VA 22904, USA}\\
\IEEEauthorblockA{\IEEEauthorrefmark{2}Hewlett Packard Labs, Hewlett Packard Enterprise, Milpitas, CA 95035, USA}\\
\IEEEauthorblockA{\IEEEauthorrefmark{3}Beijing National Research Center for Information Science and Technology, Tsinghua University, 100084, Beijing, China}\\
\IEEEauthorblockA{\IEEEauthorrefmark{4}Synopsys, Sunnyvale, CA 94085, USA}\\
\IEEEauthorblockA{\IEEEauthorrefmark{5}Dept. of Physics, University of Virginia, Charlottesville, VA 22904, USA}\\
}

%
%

\markboth{}%
{Shell \MakeLowercase{\textit{et al.}}: Bare Demo of IEEEtran.cls for IEEE Journals}
%



\maketitle

\begin{abstract}
III-V material based digital alloy Avalanche Photodiodes (APDs) have recently been found to exhibit low noise similar to Silicon APDs. The III-V  materials can be chosen to operate at any wavelength in the infrared spectrum. In this work, we present a physics-based SPICE compatible compact model for APDs built from parameters extracted from an Environment-Dependent Tight Binding (EDTB) model calibrated to {\it{ab-initio}} Density Functional Theory (DFT) and Monte Carlo (MC) methods. Using this approach, we can accurately capture the physical characteristics of these APDs in integrated photonics circuit simulations.
\end{abstract}


%
\IEEEpeerreviewmaketitle

\section{Introduction}
The rapid growth of Internet of Things (IoT) applications is resulting in the connection of more and more devices to the internet. There will be around 20.4 billion estimated IoT devices connected through machine-to-machine technology by the end of the year 2020 \cite{li20185g}. Furthermore, the advent of 5G communication technology will enable faster communication between wireless devices along with a reduction of over 90\% in energy consumption compared to 4G systems \cite{Chowdhury5G}. Thus, it is expected that 5G technology will enable an exponential growth in IoT devices and systems in the near future. This boom in telecom and data communication applications will drive the demand for more efficient and cheaper photonic integrated circuits (PICs) \cite{Liu_III_V}. Only scalable, integrated photonic technologies can meet the huge demand coming from 5G and IoT technologies. Currently, there is a significant push to integrate III-V photonic devices onto the silicon platforms, that form the backbone of modern electronic devices. The prospect of coupling optical transmitters and receivers with the state-of-the-art complementary metal oxide semiconductor (CMOS) technology for compact IoT devices is highly enticing.

In the communication area, avalanche photodiodes (APDs) can achieve superior performance than conventional \textit{p-i-n} photodiodes. The high internal gain of APDs, which arises from impact ionization, translates into greater receiver sensitivity and a dynamic operating range with accompanying increase in loss margins \cite{personick1973receiver,smith1980receiver,forrest1985sensitivity,kasper1987multigigabit}. In addition to communications \cite{campbell2008advances}, APDs have been used in a wide range of applications 
including imaging, 
\cite{bertone2007avalanche,mitra2006adaptive} and single photon detection \cite{tosi2014low,jiang2014inp}. The key challenge with APDs is the excess noise $F(M) = \langle m^2\rangle/M^2$ due to the stochastic nature of the per primary-electron avalanche gain $m$ of the impact ionization process \cite{mcintyre1966multiplication}. The tightness of the carrier distribution is given by the McIntyre formula, whose popular form can be re-expressed as
\begin{equation}
    \displaystyle F(M) - 1 = \frac{\sigma_m^2}{\langle m\rangle^2} = \displaystyle \frac{M-1}{M} + k\frac{(M-1)^2}{M}
\label{mi}
\end{equation}
with the last quadratic term arising from secondary ionization processes by impact ionization of the minority carriers. The parameter $k$ is the ratio of the hole ionization coefficient, $\beta$, to the electron ionization coefficient, $\alpha$. Many techniques have been developed to reduce this excess noise \cite{campbell_advances}. All of these approaches focus on limiting the initiation of carrier multiplication to the carrier with the highest ionization rate. 

Recently, some digital alloy III-V APDs, in essence short periodic superlattices with periodically stacked binary constituents, have experimentally demonstrated extremely low noise currents in the short infrared wave-length spectrum \cite{zheng2018InAlAs,yuan2019AlInAsSb,JPR_AlAsSb}. The underlying mechanism seems to correlate with the opening of minigaps in select subbands and the resulting change in available bandwidth and effective mass of one of the carriers, although a thorough understanding of the physics is still evolving \cite{zheng2018InAlAs,strainInAlAs,AlInAsSb_MC}. In this paper, we propose a simple circuit model of a \textit{p-i-n} APD that is calibrated to both state-of-the art first-principles material studies as well as experimental devices. The model incorporates accurate material parameters like material effective mass and bandgap as well as fitting parameters for calibration. Our model enables accurate simulation of the circuit behavior of state-of-the-art digital alloy APDs in PICs, all the way from first-principles studies of underlying materials to circuits. 

We study the material bandstructure of these alloys using an environment-dependent tight binding model (EDTB), which is calibrated to state-of-the-art DFT bandstructure and wavefunctions computed using hybrid functionals (HSE06) \cite{TanETB}. The bandstructure of some of the digital alloys with low noise characteristics show small gaps called 'minigaps' in their valence bands, that reduce hole ionization, generating a small ratio $k$ in Eq.~\ref{mi}. The transport properties of these digital alloy APDs are calculated using full-band Monte Carlo simulations, which show a good match with experimental results. In order to study their performance in circuits, a SPICE compatible model is required.  Circuit models for both \textit{p-i-n} APDs and separate absorption, charge, and multiplication (SACM) APDs have been reported \cite{chen1996pin,jou2002time,banoushi2005circuit}. Previously reported \textit{p-i-n} APD circuit models \cite{chen1996pin} use bulk material parameters that fail to capture the quantum effects, such as minigaps, seen in these short-period superlattices. Thus, it is necessary to develop a simple physics-based circuit model that can include the interesting physical properties observed in today's newer materials.   

The various tools and models used for the simulation of APDs in this paper are described in the following sections. In Section II of this paper, we describe the bandstructure model, the full-band Monte Carlo model and then the  proposed circuit model. In Section III we show simulations performed using this circuit model along with calibrations to experimental data.

\section{Model}
\begin{figure}[b]
\centering
\includegraphics[width=0.45\textwidth]{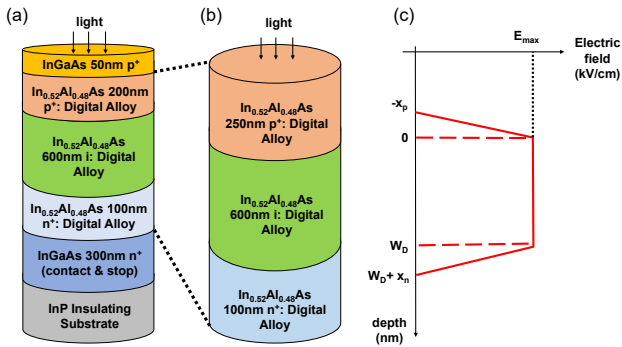}
\caption{(a) Schematic diagram of experimental InAlAs digital alloy APD (b) Schematic diagram of simplified device considered for SPICE model (c) Electric field profile of the simulated \textit{p-i-n} APD.}\label{fig:structure}
\end{figure}

In this work we consider short period III-V digital alloys for our simulation. In particular, we will study the characteristics of a digital alloy InAlAs \textit{p-i-n} APD. Fig. \ref{fig:structure}(a) shows a schematic cross-section of the device \cite{zheng2018InAlAs}. For SPICE modeling, we consider a simplified structure shown in Fig. \ref{fig:structure}(b). The typical electric field profile of this device is given in Fig. \ref{fig:structure}(c). We can see that the highest electric field is in the intrinsic region where the avalanche multiplication occurs.

The properties of III-V digital alloy APDs differ largely from their bulk counterparts  due to the band unfolding imposed by the overall superlattice periodicity. It is therefore imperative to develop a solid modeling framework that accounts for the band modification, device geometry and various scattering processes in our SPICE model. We use state-of-the-art band structure and transport models calibrated to first-principles based results and experiments, which makes our tools very reliable for simulating existing and emerging APD structures. The tools are described in the following subsections.

\subsection{Bandstructure Model}
\begin{figure}[t]
\centering
\includegraphics[width=0.48\textwidth]{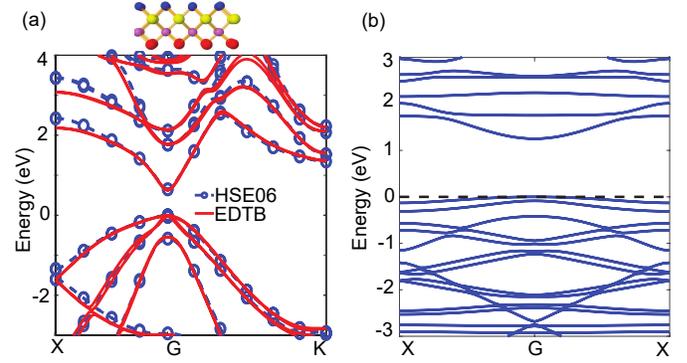}\caption{(a) Comparison of our environment-dependent tight binding model to density functional theory simulations and previous tight binding results for a InGaAsSb alloy \cite{AhmedTFET}.(b) Bandstructure of 6-monolayer digital alloy InAlAs considered in this paper.}\label{fig:bandstructure}
\end{figure}

As a first step, the detailed band structure of the material is calculated using the environment dependent tight binding (EDTB) model. The accuracy of the EDTB is underscored by the need for precise material parameters used in the circuit simulations. Conventional tight binding is benchmarked to bulk bandstructures and not readily transferrable to surfaces and interfaces where the environment impacts the chemistry substantially. Readjusting the parameters is not straightforward, as these methods work directly with the eigenvalues, i.e., E-k relationships, and do not work with the full eigenvectors, especially their radial parts that determine bonding and tunneling (only the angular parts of the orbitals are used to enforce the overall group theoretical symmetry). 
Our past attempt at due diligence on radial eigenvectors was to use Extended H\"uckel theory \cite{huckel_cnt,huckel_silicon} which employed explicit Wannier basis sets constructed from non-orthogonal atomic orbitals that were fitted to arrive at the bulk Hamiltonian. Since the basis sets were fitted rather than the energies, the latter were transferrable to other environments that simply required recomputing the overlap matrix elements. In contrast,  the EDTB model works within conventional orthogonal Wannier like basis sets, and is not only calibrated to DFT band structure but also  wavefunctions that are based on experiments and first-principles calculations \cite{TanETB,tan2015tight}. For the DFT bandstructure and wavefunctions calculations we employ the HSE06 hybrid functional approximation  which predicts material band structure accurately \cite{heyd2003hybrid}. The accuracy of our EDTB model for a InGaAsSb alloy is described in Fig. \ref{fig:bandstructure}(a) \cite{AhmedTFET}. It shows that the bandstructure of InGaAsSb calculated with the EDTB model replicates the bandstructure obtained using DFT very precisely. Fig. \ref{fig:bandstructure}(b) shows the EDTB band structure of a 6-monolayer InAlAs digital alloy used in this simulation. The resulting band structure is fed into a full band Monte Carlo simulator. 

\subsection{Monte Carlo Simulation} \label{monte_carlo}

\begin{figure}[t]
\centering
\includegraphics[width=0.48\textwidth]{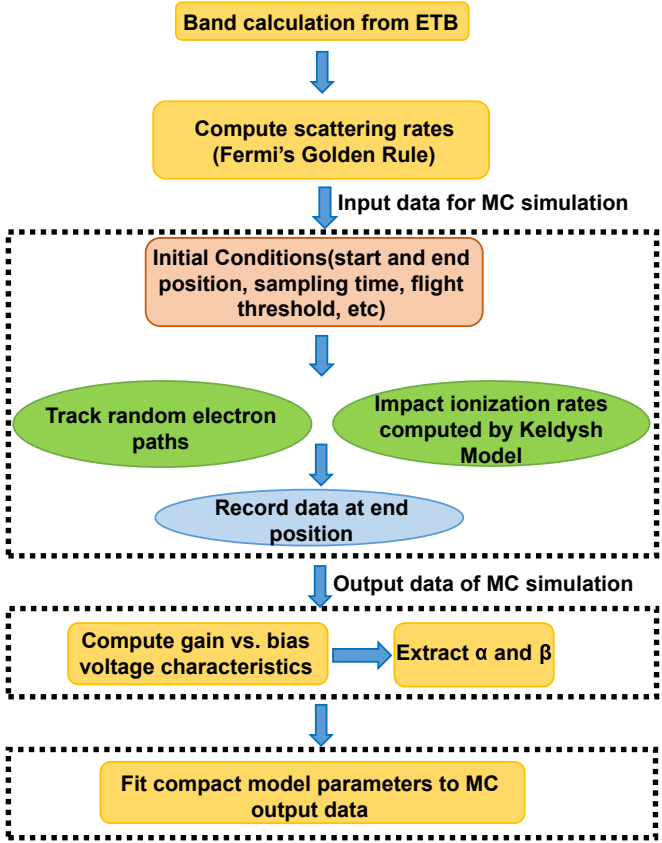}
\caption{Flowchart of Monte Carlo simulation process.}\label{fig:monte carlo}
\end{figure}

The Monte Carlo simulation tracks the transport behavior of injected electrons. The flowchart of the method used for extracting device parameters is shown in Fig. \ref{fig:monte carlo}. Initially, the scattering rates are calculated using Fermi's Golden Rule, incorporating the full bandstructure obtained from the EDTB model. The Monte Carlo simulation results used in this paper consider deformational potential scattering. The deformational scattering rate $P^{def}_{\nu \nu',\eta}(\textbf{k},\Omega _{\textbf{k}\pm \textbf{q}})$ from a point $\textbf{k}$ in band $\nu$ to a region $\Omega _{\textbf{k}'}$ in band $\nu'$ centered around $\textbf{k'}$ is expressed as \cite{InAlAs_MC,AlInAsSb_MC}

\begin{equation}
\begin{split}
P^{def}_{\nu \nu',\eta}(\textbf{k},\Omega _{\textbf{k}\pm \textbf{q}})&=\frac{\pi}{\rho \omega_{nq}} | \Delta^{\eta} (\nu',\textbf{k},\textbf{q},\nu)|^2 |I(\nu,\nu';\textbf{k},\textbf{k} \pm \textbf{q})|^2 \\
    & D_{\nu'}(E',\Omega _{\textbf{k}'} \left(N_q+\frac{1}{2} \mp \frac{1}{2}\right)\\
\end{split}
\end{equation}

 where, $\rho$ represents the lattice density, $\textbf{q}$ is the phonon wave vector of mode $\eta$ and  the deformation potential is $\Delta^{\eta} (\nu',\textbf{k},\textbf{q},\nu)$.

The path of a single electron through the multiplication region is then tracked under the effects of electric field and random scattering events. The impact ionization rates used in the simulation are computed using the Keldysh model \cite{dunn1997monte},  described below. The impact ionization rate using this model is expressed by \cite{AlInAsSb_MC,InAlAs_MC}

\begin{equation}
    P_{\nu \nu'}(\textbf{k},\textbf{k}')=S(E-E_{th})^\gamma
\end{equation}

where, $E_{th}$ is the threshold energy, $S$ is the softness parameter and $\gamma$ is an approaching index. We can approximate the threshold energy by $E_{TH}=E_{G} (2\mu^{-1}+1)/(\mu^{-1}+1) $ where $E_G$ is the material bandgap and $\mu^{-1}$ is the ratio of the electron effective mass to the hole effective mass. These parameters were adjusted by fitting the gain curves and excess noise of the Monte Carlo simulation results with experimental results. The simulation is repeated for multiple electrons and carrier transport properties are computed by averaging over the many trajectories \cite{zheng2016pmt}. From the ensemble Monte Carlo simulation, we obtain the gain versus bias voltage characteristics, as shown in Fig. \ref{fig:gain} for an InAlAs APD.

The electron and hole ionization coefficients, $\alpha$ and $\beta$ respectively, can be calculated using the following equations: 

\begin{figure}[b]
\centering
\includegraphics[width=0.3\textwidth]{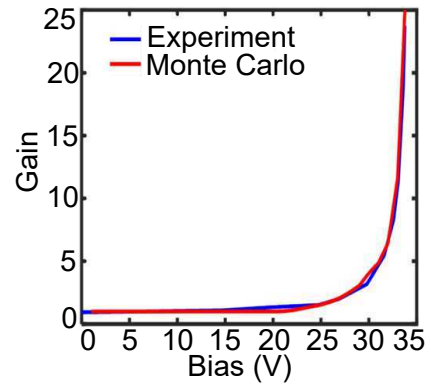}
\caption{Gain vs. reverse bias voltage characteristics for the InAlAs APD.}\label{fig:gain}
\end{figure}

\begin{equation}
    \alpha({\cal{E}})=\frac{1}{W_D}\left[\frac{M_n(V)-1}{M_n(V)-M_p(V)}\right]ln\left[\frac{M_n(V)}{M_p(V)}\right]
\end{equation}

\begin{equation}
    \beta({\cal{E}})=\frac{1}{W_D}\left[\frac{M_p(V)-1}{M_p(V)-M_n(V)}\right]ln\left[\frac{M_p(V)}{M_n(V)}\right]
\end{equation}
where, $M_n(V)$ and $M_p(V)$ represent the gain of the electrons and holes respectively, ${\cal{E} = V/W_D}$ is the electric field and $W_D$ is the width of the multiplication region. This ratio is an important metric for determining the excess noise factor $F(M)$ of APDs. By varying parameters like temperature and repeating the MC simulations we can extract the relationship of the ionization coefficients with these parameters. Then the extracted ionization coefficients, their ratio and the gain can be re-expressed in terms of simpler empirical functions that are then used for the compact model:
\begin{equation}
   k=c_1 e^{c_2T}
   \label{k_temp}
\end{equation}

\begin{equation}
    \alpha(T,{\cal{E}})=c_3 exp\left[-c_{4}T-\left(\frac{c_5}{{\cal{E}}}\right)^{\displaystyle n}\right]
    \label{alpha_temp}
\end{equation}

\begin{equation}
    M=\frac{k-1}{k-exp\left[\alpha(1-k)W_D\right]}
    \label{gain_temp}
\end{equation}

\begin{center}
\begin{figure*}[t]
\centering
\includegraphics[width=18cm,height=9cm]{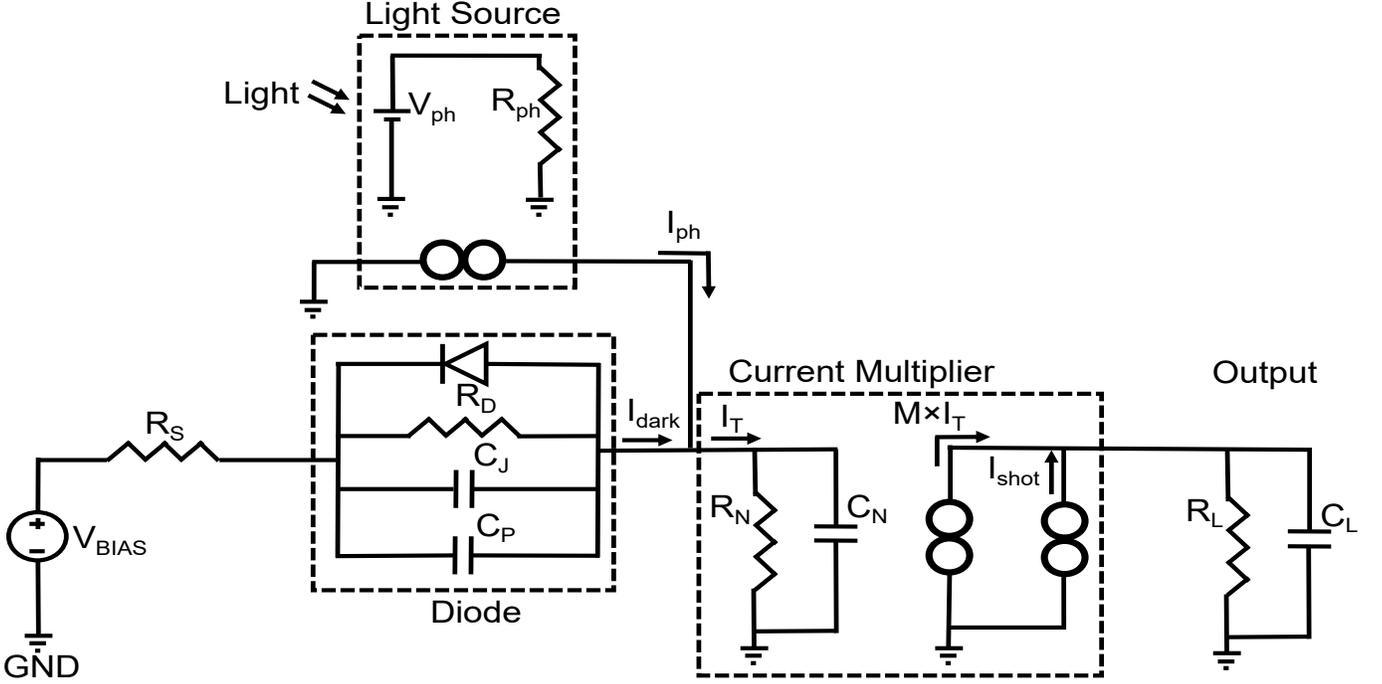}
\caption{Schematic diagram of avalanche photodiode model and testbench used in the SPICE simulations.}\label{fig:schematic}
\end{figure*}
\end{center}
where $M = M_n$ for pure electron injection and $M_p$ for pure hole injection. In other words, the ionization coefficients are reduced to six material-dependent constants $c_{1-5}$ and $n$ that are extracted from numerical data on the gain curve, such as from experiments or the ensemble Monte Carlo approach described earlier. The fitted parameters are used to calculate the voltage dependent impact ionization gain in the circuit model.

\subsection{Circuit Model} \label{circuit_model}

\begin{table}[b]
\centering
\begin{tabular}[h]{ |c|c| }  
 \hline
 \textbf{Parameter} & \textbf{Value} \\
 \hline
 $\eta$ & 40\% \\  
 \hline
 $c_1$ & 0.012 \\
 \hline
 $R$ & 0.01 \\ 
 \hline
 $c_2$ & 0.0147 $K^{-1}$ \\
 \hline
  $m^*$ & 0.08$m_0$ \\  
  \hline
  $c_3$ & $2.2\times 10^7$ $cm^{-1}$ \\ 
  \hline
   $E_G$ & 1.25 $eV$ \\
   \hline
   $c_4$ & 0.004 $K^{-1}$ \\
   \hline
    $n$ & 0.9 \\
    \hline
    $c_5$ & $3.5\times 10^6$ $V cm^{-1}$ \\
\hline
$a_p$ & $1.57\times10^4$ $V cm^{-1}$ \\ 
\hline
$A$ & 31.4 $nm^2$ \\
\hline
$\lambda$ & 1.08 $\mu m$ \\
\hline
$R_D$ & $1.5\times 10^{11}$ $\Omega$ \\
\hline
$I_{L0}$ & $5.3\times10^{-13}$ $A$ \\
\hline 
$\zeta$ & 0.3414 $V^{-1}$ \\
\hline
$k$ & 0.01 \\
 \hline
 $\Theta$ & 0.8 \\
 \hline
  T & 300 $K$ \\
 \hline
   $W_P$ & 250 $nm$ \\
 \hline
   $C_J$ & 5.77 $pF$ \\
 \hline
   $R_L$ & 0.1 $\Omega$ \\
 \hline
  $C_L$ & 1 $pF$ \\
 \hline
\end{tabular}
\caption{Table of major parameter values used in SPICE model}
\label{table:1}
\end{table} 

In this section we describe the SPICE compatible physics-based circuit model. The circuit layout of the compact model is displayed in Fig. \ref{fig:schematic}. The unit cell consists of a simple diode coupled to a light source unit and a current multiplier unit to model the APD characteristics. For our model, we consider a dark current that consists of two components: the reverse bias saturation current and the tunneling current. The dominating current at low reverse bias is the current through the resistor $R_D$. The value of $R_D$ can be calibrated from experimental $I-V$ characteristics or theoretically computed by diode reverse bias saturation current using material parameters. The tunneling current flowing through the diode symbol, $I_{TUN}$, considering a triangular barrier, is calculated using the Fowler-Nordheim equation.

\begin{equation}
    I_{TUN}=\frac{\sqrt{2m^*}q^3{\cal{E}}V_{bias}A}{4\pi^2\hbar^2 E_G^{1/2}} exp\left[-\frac{\Theta \sqrt{2m^*} E_G^{3/2}}{q\hbar {\cal{E}}}\right]
\end{equation}
The parameters $m^*$, $V_{bias}$, $A$ and $E_G$ represent the tunneling effective mass, the bias voltage across the \textit{p-i-n} structure, cross-sectional area and the bandgap, respectively. The fitting parameter $\Theta$ is used to account for the difference in barrier shape with that of an experimental device. For a triangular barrier we can consider $\Theta=4/3$.  Another current component, $I_L$,       is used to model other various leakage currents through the diode like Shockley-Read-Hall generation and trap-assisted tunneling 
\begin{equation}
    I_L=I_{L0}e^{\zeta V_{bias}}
\end{equation}
For calibration with InAlAs APD experimental data and simulations in this paper we consider $I_{L0}$ and $\zeta$ as fitting parameters. The total dark current, $I_{dark}$, is then given by

\begin{equation}
    I_{dark}=\frac{V_{bias}}{R_D}+I_{TUN}+I_L
\end{equation}

The photocurrent $I_{ph}$ is modeled as a voltage controlled current source.  The photon source is modeled as a voltage source, $V_{ph}$, connected to a large resistor, $R_{ph}$. The source input power $P_{IN}$, which is proportional to ${V}_{ph}$, is determined by 
\begin{equation}
    P_{IN}=V_{ph}\times 1amp
    \label{photon_input}
\end{equation}

\begin{equation}
    I_{ph}=q\eta\frac{P_{IN}(1-R)}{hc/\lambda}\left[1-e^{\displaystyle -a_p W_p}\right]
    \label{photocurrent}
\end{equation}
Eq.\ref{photocurrent} gives the photocurrent, $I_{ph}$. In the equation $R$ is the reflectivity of the absorption region, $a_p$ is the absorption coefficient, $W_P$ is the \textit{p}-region width and $\lambda$ gives the wavelength of the input light source. $\eta$ represents the internal quantum efficiency. For a given material, the theoretical value of $\eta$ can be numerically calculated using \cite{GangulyPbSe}

\begin{equation}
    \eta\left(\lambda,T\right)=\frac{E_g \int_{E_G}^\infty D(E) f(E) dE}{\int_{0}^\infty E D(E) f(E) dE}
\end{equation}
where, $D(E)$ and $f(E)$ are the photon density of states and occupancy (given by Bose-Einstein statistics) functions.

The reason a voltage source is preferred for the photon source is that internally SPICE models use all nodal equations as $I = GV$ type matrices when the simulations are set up, i.e. the voltages $V$ are independent variables, and the currents $I$ are dependent. Therefore, any current-controlled voltage source or current-controlled current source element needs to be converted to an equivalent voltage-controlled current source or voltage-controlled voltage source element using Thevenin equivalent. This matters only for large simulations or to guard against convergence issues, particularly if the resistance, $R$, is very small since this makes the conductance  $G$ matrix singular. In this case since $R$ is very large this is not a consideration.

\begin{figure}[b]
\centering
\includegraphics[width=0.48\textwidth]{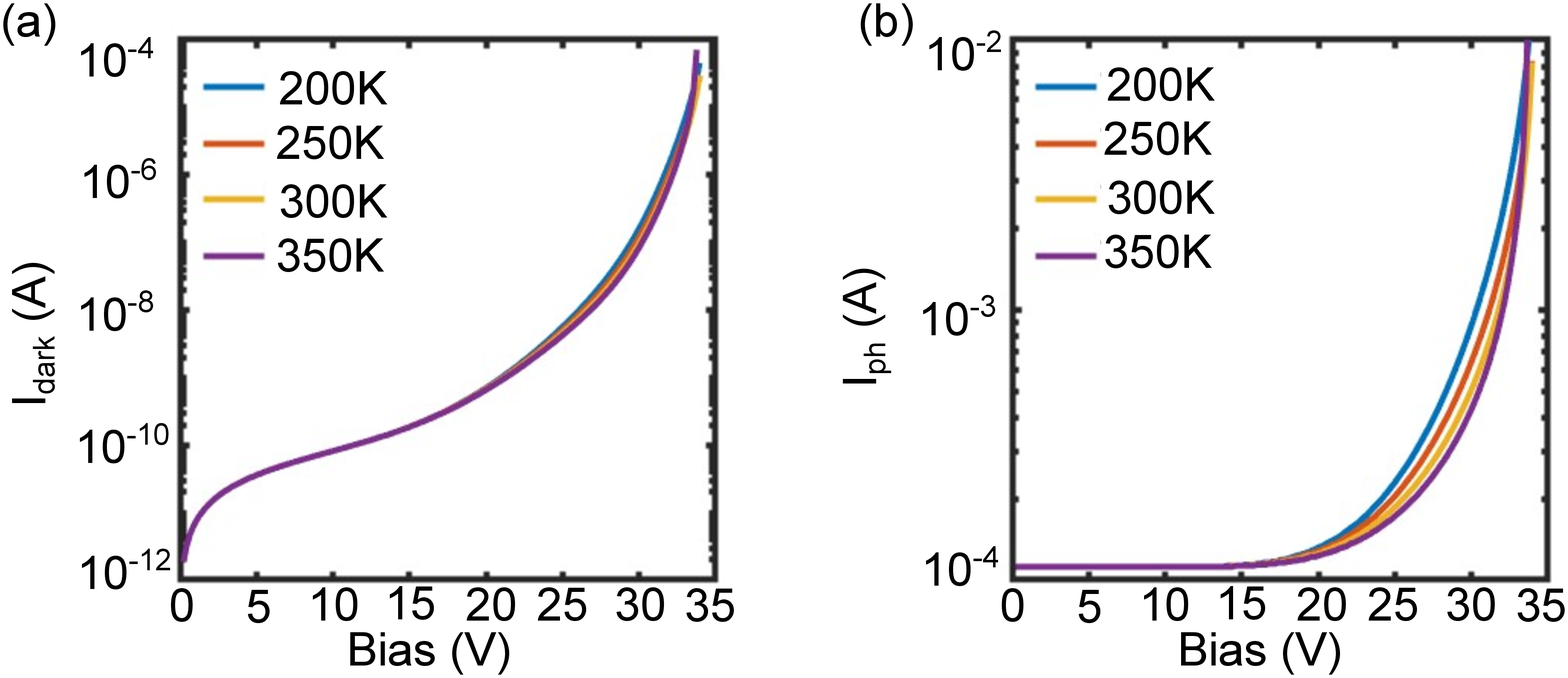}
\caption{(a) Dark current vs. reverse bias voltage characteristics (b) Photo current vs. reverse bias voltage characteristics. Both plots are given as a function of temperature.}\label{fig:IV}
\end{figure}

The current multiplier models the avalanche multiplication of the APD. The multiplier takes the sum of the unmultiplied dark and photo current given by eq. \ref{unmultiplied current} as input and multiplies that by the gain $M$, which is a function of the reverse bias voltage given by Eq. \ref{gain_temp}. The output current of the APD is finally obtained by eq. \ref{output current} 
\begin{equation}
    I_T=I_{dark}+I_{ph}
    \label{unmultiplied current}
\end{equation}

\begin{equation}
    I_{out}=M\times I_T
    \label{output current}
\end{equation}
 
The noise sources of the APDs must also be included in the model to accurately model their behavior in circuit applications. In the model, noise currents are modeled as separate current sources. In this paper, we can consider only the dominant shot noise of the APDs. For a given bandwidth $\Delta f$, the noise variance of the shot noise is given by

\begin{equation}
 \langle i^2_{shot}\rangle =2q(I_{dark}+I_{ph})\langle M^2\rangle F(M)\Delta f   
\end{equation}
 
It is possible to extract the APD bandwidth, $\Delta f$, from MC simulation. The average impulse function for a large number of short input photon pulses, shaped like Gaussian pulses, can be extracted from MC simulation \cite{huntington2020ingaas}. The decaying tail of the the impulse function can be fitted by the function $h(t)=2\pi\eta Mq \Delta f exp(-2\pi \Delta f t) u(t)$, where $u(t)$ is a unit step function, having a value of one for all positive t and zero for the rest. Here, $\Delta f$ is the frequency at which APD's gain drops to 0.707 of its maximum value. It is important to point out that the bandwidth of a circuit, where APD is a component, can be dominated by other circuit elements besides the APD. However, those extrinsic effects can be captured by our compact circuit model where the APD enters as a component block.

For transient calculations, capacitances are added in parallel to the diode. $C_J$ represents the intrinsic capacitance of the device and $C_P$ is the parasitic capacitance. A resistor $R_S$ is added in series to the bias voltage source to model the contact resistances. A small resistor and capacitor, $R_N$ and $C_N$, are used at the output of the diode to help with convergence in SPICE. $R_L$ and $C_L$ are the load resistor and capacitor used to extract the output. The values used in this paper are listed in Table I and are obtained from a previous publication \cite{yuan2018temperature}. These can be computed from MC simulations.

In summary, given any digital alloy combination our compact model can take material parameters like bandgap and effective mass as input and outputs relevant metrics of APD circuit simulation like dark current, gain, quantum efficiency and noise current. The dark current is obtained from eq, (11), gain from eq. (8), quantum efficiency from eq. (14) and APD excess noise current from eq. (17). These metrics are calculated based on calibrations to  MC simulation or experimental results.

\section{Results and Discussion}
The temperature dependent reverse bias I-V characteristics are shown in Fig. \ref{fig:IV}(a). At low bias the current is dominated by the diffusion current density which is a constant in the device. The large current at high bias is attributed to the increasing tunneling current and increasing gain of the device. At high bias, the current reduces with increased temperature due to increased phonon scattering occurring in the device, captured through our scattering rates that enter the Monte Carlo simulations and are eventually lumped into the $c_{1-5}$ and $n$ parameters. Fig. \ref{fig:IV}(b) shows the photo-current vs. reverse bias voltage for different temperatures. It exhibits a similar trend at high bias like the dark current due to the increasing gain and phonon scattering.  For this plot we considered $P_{IN} =1mW$.

\begin{figure}[t]
\centering
\includegraphics[width=0.48\textwidth]{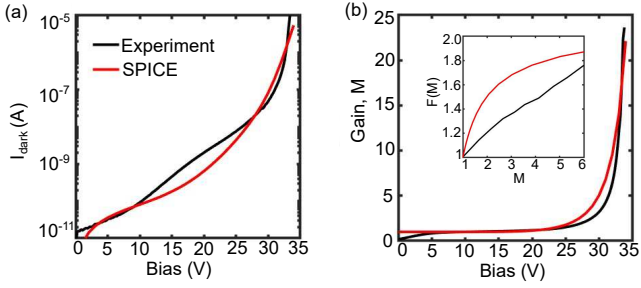}
\caption{(a) Calibration of the simulated dark current characteristics using SPICE model of a 6-monolayer InAlAs digital alloy APD (b) Calibrated of the simulated gain vs. bias characteristics of this APD. Inset-Excess noise factor vs. gain of computed using McIntyre’s Formula compared to the experimental results.}\label{fig:expt_IV}
\end{figure}

\begin{figure}[b]
\centering
\includegraphics[width=0.3\textwidth]{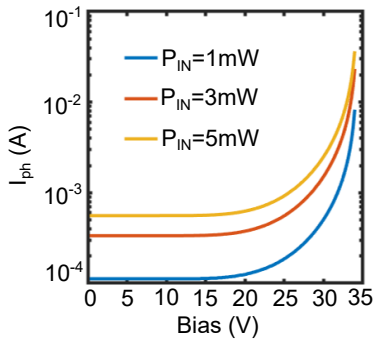}
\caption{Photocurrent vs. reverse bias voltage characteristics as a function of input power.}\label{fig:Pin_IV}
\end{figure}

In Fig. \ref{fig:expt_IV}(a) the simulated reverse bias dark current characteristics of the InAlAs digital  alloy APD using the calibrated compact model are plotted and compared to the experimental measurements \cite{zheng2018InAlAs}. The calibrated parameters are given in Table \ref{table:1}. The simulated values agree well at low and high bias. There is a small discrepancy in the middle region that is due to the use of the approximate equations used in calculating some of the parameters which results in overestimation of the scattering processes involved. Additionally, some discrepancy is due to the simplification of the structure considered in the simulation. This is primarily to allow for faster convergence of the SPICE model. Fig. \ref{fig:expt_IV}(b) shows the simulated gain versus bias characteristics compared to the experimental data. The values are in good agreement with each other. To calculate the excess noise factor versus gain characteristics we use the standard McIntyre’s formula \cite{mcintyre1966multiplication}. The comparison between the calculated and experimental F versus M is shown in the inset of Fig. \ref{fig:expt_IV}(b). We observe that McIntyre's formula overestimates the excess noise at low gain values compared to experiments. The mismatch can be attributed to simplifications, like ignoring the `dead-space' effect.

\begin{figure}[t]
\centering
\includegraphics[width=0.48\textwidth]{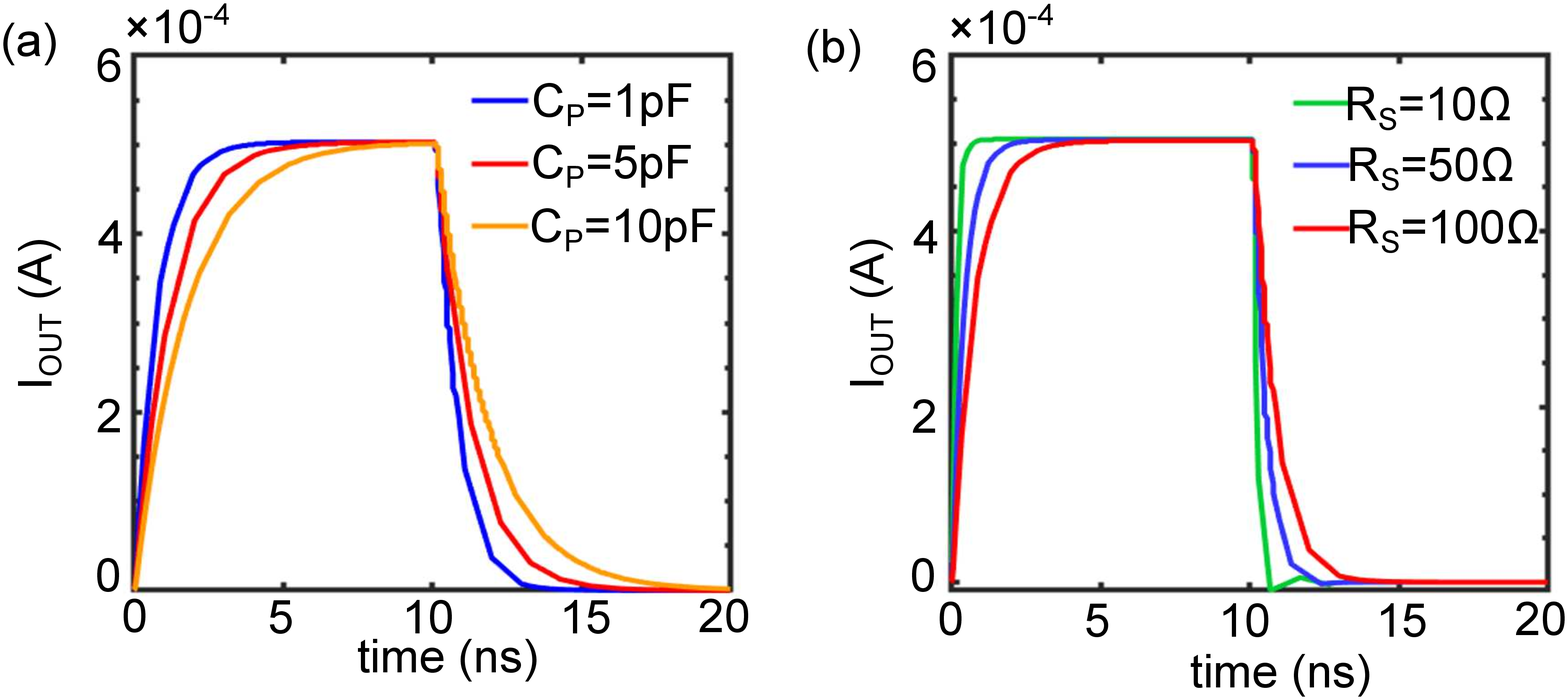}
\caption{(a) Transient response of output current with varying parasitic capacitance, $C_P$ (b) Transient response of output current as a function of contact resistance, $R_S$.}\label{fig:Cp_Rs}
\end{figure}

We plot the output I-V characteristics for different input powers in Fig. \ref{fig:Pin_IV}. A higher input power of the light source results in more electron-hole pair generation in the absorption region. This increases the current at low bias. At very high bias, the APD reaches the avalanche breakdown region where the output current is not affected much by the input power magnitude due to the high carrier generation by the impact ionization process.

\begin{figure}[b]
\centering
\includegraphics[width=8.75cm, height=5.5cm]{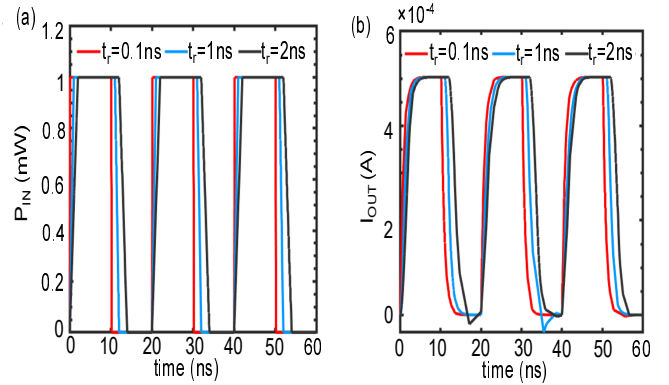}
\caption{a) Input power (light source) vs. time for different rise and fall times (b) output current vs. time for the different rise and fall times.}\label{fig:transient simulation}
\end{figure}

The transient characteristics of the InAlAs digital alloy APD considered in this compact model are shown in Figs. \ref{fig:Cp_Rs} and \ref{fig:transient simulation}. The bias voltage used is 30V. The transient response of the output current for different parasitic capacitance is simulated in Fig. \ref{fig:Cp_Rs}(a). We can see that the parasitic capacitance present in the APD can significantly affect its responsiveness to the input light. A larger capacitance results in longer rise and fall times. In a receiver this will translate to lower detection speed and will affect the overall speed of an integrated photonic device.  The effect of different contact resistances on the output transient response is shown in Fig. \ref{fig:Cp_Rs}(b). Increased contact resistance also leads to higher rise and fall times. However, the effect of $R_S$ is not as significant as the parasitic capacitances. 

Fig. \ref{fig:transient simulation} shows input and output characteristics of the APD for different rise and fall times. It is seen that a longer rise/fall time results in slower discharge at the end of each input cycle and can also result in overshoots while discharging. This shows that switching pattern of the light source can also affect the refresh rate of the APD.

\begin{figure}[t]
\centering
\includegraphics[width=0.3\textwidth]{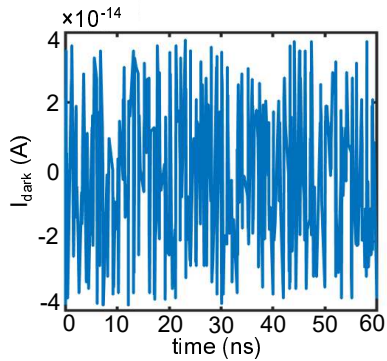}
\caption{Simulated shot noise of the InAlAs APD.}\label{fig:noise}
\end{figure}

One of the crucial elements of APD performance is the excess noise that is generated due to the random nature of the impact ionization process. Any noise can significantly degrade the photon detection ability of these photodetectors. Thus, it is necessary to include the noise in the circuit model in order to accurately model PICs. This noise is reduced by allowing impact ionization to be initiated by carrier injection of the carrier with the highest ionization coefficient. This essentially translates to a lower value of the ionization coefficient ratio $k$. 

Fig. \ref{fig:noise} shows the dominant shot noise in the dark current simulated using the SPICE model. The noise current is added as a separate current source in the circuit. We consider $\Delta f=1Hz$ since it is generally application dependent. Many methods are used to reduce the shot noise in the APDs. In the InAlAs digital alloy APD the noise is minimized due to the presence of “minigaps” in the valence band which prevents holes from ionizing and thus suppress $F(M)$ \cite{zheng2018InAlAs}. Although in this treatment we only consider shot noise as the dominant source in digital APDs, we can in principle include other types of noise relevant for different geometries as added current sources.

Despite the model being developed for \textit{p-i-n} APDs, it can be modified to simulate othe structures such as Separate Absorption Charge Multiplication (SACM) APDs as well. The potential profile of the SACM APDs can be obtained using added electrostatic solvers, which can then be used as input to the MC model. The fitting parameters of the compact model, described in sections \ref{monte_carlo} and \ref{circuit_model}, can be adjusted to match the results of the MC simulations. The compact model might require some additional resistance and capacitance elements to account for the behavior of the absorption and charge layers of the SACM structure.

\section{Conclusion}
In this paper, we report an elementary physics-based compact model of APDs. This model captures the essential device physics calculated by first-principles methods and simulates their effect on circuit behavior. The compact model ultimately treats the APD as a black box that can be included in a modular fashion into bigger photonic integrated circuits, with its input being material and geometric parameters and output being relevant performance metrics such as excess noise, gain, dark current and quantum efficiency. In particular, we simulate the behavior of a new class of digital alloy APDs which exhibit low excess noise at very high bias, which makes them highly suitable for a wide range of applications.


%

\section*{Acknowledgment}
This work was funded by National Science Foundation grant NSF 1936016. We acknowledge fruitful discussions with Seth Banks and Andreas Beling.

\ifCLASSOPTIONcaptionsoff
  \newpage
\fi



%

\bibliographystyle{IEEEtran}
\bibliography{IEEEabrv,IEEEexample.bib}

\begin{thebibliography}{10}
\providecommand{\url}[1]{#1}
\csname url@samestyle\endcsname
\providecommand{\newblock}{\relax}
\providecommand{\bibinfo}[2]{#2}
\providecommand{\BIBentrySTDinterwordspacing}{\spaceskip=0pt\relax}
\providecommand{\BIBentryALTinterwordstretchfactor}{4}
\providecommand{\BIBentryALTinterwordspacing}{\spaceskip=\fontdimen2\font plus
\BIBentryALTinterwordstretchfactor\fontdimen3\font minus
  \fontdimen4\font\relax}
\providecommand{\BIBforeignlanguage}[2]{{%
\expandafter\ifx\csname l@#1\endcsname\relax
\typeout{** WARNING: IEEEtran.bst: No hyphenation pattern has been}%
\typeout{** loaded for the language `#1'. Using the pattern for}%
\typeout{** the default language instead.}%
\else
\language=\csname l@#1\endcsname
\fi
#2}}
\providecommand{\BIBdecl}{\relax}
\BIBdecl

\bibitem{li20185g}
S.~Li, L.~Da~Xu, and S.~Zhao, ``5g internet of things: A survey,''
  \emph{Journal of Industrial Information Integration}, vol.~10, pp. 1--9,
  2018.

\bibitem{Chowdhury5G}
M.~Z. {Chowdhury}, M.~K. {Hasan}, M.~{Shahjalal}, E.~B. {Shin}, and Y.~M.
  {Jang}, ``Opportunities of optical spectrum for future wireless
  communications,'' in \emph{2019 International Conference on Artificial
  Intelligence in Information and Communication (ICAIIC)}, Feb 2019, pp.
  004--007.

\bibitem{Liu_III_V}
A.~Y. {Liu} and J.~{Bowers}, ``Photonic integration with epitaxial iii–v on
  silicon,'' \emph{IEEE Journal of Selected Topics in Quantum Electronics},
  vol.~24, no.~6, pp. 1--12, Nov 2018.

\bibitem{personick1973receiver}
S.~D. Personick, ``Receiver design for digital fiber optic communication
  systems, parts i and ii,'' \emph{Bell system technical journal}, vol.~52,
  no.~6, pp. 843--886, 1973.

\bibitem{smith1980receiver}
R.~Smith and S.~Personick, ``Receiver design for optical fiber communication
  systems,'' in \emph{Semiconductor devices for optical communication}.\hskip
  1em plus 0.5em minus 0.4em\relax Springer, 1980, pp. 89--160.

\bibitem{forrest1985sensitivity}
S.~Forrest, ``Sensitivity of avalanche photodetector receivers for
  high-bit-rate long-wavelength optical communication systems,'' in
  \emph{Semiconductors and Semimetals}.\hskip 1em plus 0.5em minus 0.4em\relax
  Elsevier, 1985, vol.~22, pp. 329--387.

\bibitem{kasper1987multigigabit}
B.~Kasper and J.~Campbell, ``Multigigabit-per-second avalanche photodiode
  lightwave receivers,'' \emph{Journal of lightwave technology}, vol.~5,
  no.~10, pp. 1351--1364, 1987.

\bibitem{campbell2008advances}
J.~C. Campbell, ``Advances in photodetectors,'' in \emph{Optical Fiber
  Telecommunications VA}.\hskip 1em plus 0.5em minus 0.4em\relax Elsevier,
  2008, pp. 221--268.

\bibitem{bertone2007avalanche}
N.~Bertone and W.~Clark, ``Avalanche photodiode arrays provide versatility in
  ultrasensitive applications,'' \emph{Laser focus world}, vol.~43, no.~9,
  2007.

\bibitem{mitra2006adaptive}
P.~Mitra, J.~Beck, M.~Skokan, J.~Robinson, J.~Antoszewski, K.~Winchester,
  A.~Keating, T.~Nguyen, K.~Silva, C.~Musca \emph{et~al.}, ``Adaptive focal
  plane array (afpa) technologies for integrated infrared microsystems,'' in
  \emph{Intelligent Integrated Microsystems}, vol. 6232.\hskip 1em plus 0.5em
  minus 0.4em\relax International Society for Optics and Photonics, 2006, p.
  62320G.

\bibitem{tosi2014low}
A.~Tosi, N.~Calandri, M.~Sanzaro, and F.~Acerbi, ``Low-noise, low-jitter, high
  detection efficiency ingaas/inp single-photon avalanche diode,'' \emph{IEEE
  Journal of selected topics in quantum electronics}, vol.~20, no.~6, pp.
  192--197, 2014.

\bibitem{jiang2014inp}
X.~Jiang, M.~Itzler, K.~O’Donnell, M.~Entwistle, M.~Owens, K.~Slomkowski, and
  S.~Rangwala, ``Inp-based single-photon detectors and geiger-mode apd arrays
  for quantum communications applications,'' \emph{IEEE Journal of Selected
  Topics in Quantum Electronics}, vol.~21, no.~3, pp. 5--16, 2014.

\bibitem{mcintyre1966multiplication}
R.~McIntyre, ``Multiplication noise in uniform avalanche diodes,'' \emph{IEEE
  Transactions on Electron Devices}, no.~1, pp. 164--168, 1966.

\bibitem{campbell_advances}
J.~C. {Campbell}, ``Recent advances in avalanche photodiodes,'' \emph{Journal
  of Lightwave Technology}, vol.~34, no.~2, pp. 278--285, Jan 2016.

\bibitem{zheng2018InAlAs}
J.~Zheng, Y.~Yuan, Y.~Tan, Y.~Peng, A.~K. Rockwell, S.~R. Bank, A.~W. Ghosh,
  and J.~C. Campbell, ``Digital alloy inalas avalanche photodiodes,''
  \emph{Journal of Lightwave Technology}, vol.~36, no.~17, pp. 3580--3585,
  2018.

\bibitem{yuan2019AlInAsSb}
Y.~{Yuan}, A.~K. {Rockwell}, Y.~{Peng}, J.~{Zheng}, S.~D. {March}, A.~H.
  {Jones}, M.~{Ren}, S.~R. {Bank}, and J.~C. {Campbell}, ``Comparison of
  different period digital alloy al${}_{\text{0.7}}$inassb avalanche
  photodiodes,'' \emph{Journal of Lightwave Technology}, vol.~37, no.~14, pp.
  3647--3654, July 2019.

\bibitem{JPR_AlAsSb}
\BIBentryALTinterwordspacing
X.~Yi, S.~Xie, B.~Liang, L.~Lim, J.~Cheong, M.~Debnath, D.~Huffaker, C.~Tan,
  and J.~David, ``Extremely low excess noise and high sensitivity
  alas0.56sb0.44 avalanche photodiodes,'' \emph{Nature Photonics}, July 2019,
  {\copyright} 2019 The Authors. [Online]. Available:
  \url{http://eprints.whiterose.ac.uk/148333/}
\BIBentrySTDinterwordspacing

\bibitem{strainInAlAs}
\BIBentryALTinterwordspacing
J.~Zheng, Y.~Tan, Y.~Yuan, A.~W. Ghosh, and J.~C. Campbell, ``Strain effect on
  band structure of inalas digital alloy,'' \emph{Journal of Applied Physics},
  vol. 125, no.~8, p. 082514, 2019. [Online]. Available:
  \url{https://doi.org/10.1063/1.5045476}
\BIBentrySTDinterwordspacing

\bibitem{AlInAsSb_MC}
\BIBentryALTinterwordspacing
J.~Zheng, S.~Z. Ahmed, Y.~Yuan, A.~Jones, Y.~Tan, A.~K. Rockwell, S.~D. March,
  S.~R. Bank, A.~W. Ghosh, and J.~C. Campbell, ``Full band monte carlo
  simulation of alinassb digital alloys,'' \emph{InfoMat}, vol.~2, no.~6, pp.
  1236--1240, 2020. [Online]. Available:
  \url{https://onlinelibrary.wiley.com/doi/abs/10.1002/inf2.12112}
\BIBentrySTDinterwordspacing

\bibitem{TanETB}
\BIBentryALTinterwordspacing
Y.~Tan, M.~Povolotskyi, T.~Kubis, T.~B. Boykin, and G.~Klimeck, ``Transferable
  tight-binding model for strained group iv and iii-v materials and
  heterostructures,'' \emph{Phys. Rev. B}, vol.~94, p. 045311, Jul 2016.
  [Online]. Available:
  \url{https://link.aps.org/doi/10.1103/PhysRevB.94.045311}
\BIBentrySTDinterwordspacing

\bibitem{chen1996pin}
W.~Chen and S.~Liu, ``Pin avalanche photodiodes model for circuit simulation,''
  \emph{IEEE Journal of Quantum Electronics}, vol.~32, no.~12, pp. 2105--2111,
  1996.

\bibitem{jou2002time}
J.-J. Jou, C.-K. Liu, C.-M. Hsiao, H.-H. Lin, and H.-C. Lee, ``Time-delay
  circuit model of high-speed pin photodiodes,'' \emph{IEEE Photonics
  technology letters}, vol.~14, no.~4, pp. 525--527, 2002.

\bibitem{banoushi2005circuit}
A.~Banoushi, M.~Kardan, and M.~A. Naeini, ``A circuit model simulation for
  separate absorption, grading, charge, and multiplication avalanche
  photodiodes,'' \emph{Solid-state electronics}, vol.~49, no.~6, pp. 871--877,
  2005.

\bibitem{huckel_cnt}
\BIBentryALTinterwordspacing
D.~Kienle, J.~I. Cerda, and A.~W. Ghosh, ``Extended h\"uckel theory for band
  structure, chemistry, and transport. i. carbon nanotubes,'' \emph{Journal of
  Applied Physics}, vol. 100, no.~4, p. 043714, 2006. [Online]. Available:
  \url{https://doi.org/10.1063/1.2259818}
\BIBentrySTDinterwordspacing

\bibitem{huckel_silicon}
\BIBentryALTinterwordspacing
D.~Kienle, K.~H. Bevan, G.-C. Liang, L.~Siddiqui, J.~I. Cerda, and A.~W. Ghosh,
  ``Extended h\"uckel theory for band structure, chemistry, and transport. ii.
  silicon,'' \emph{Journal of Applied Physics}, vol. 100, no.~4, p. 043715,
  2006. [Online]. Available: \url{https://doi.org/10.1063/1.2259820}
\BIBentrySTDinterwordspacing

\bibitem{tan2015tight}
Y.~P. Tan, M.~Povolotskyi, T.~Kubis, T.~B. Boykin, and G.~Klimeck,
  ``Tight-binding analysis of si and gaas ultrathin bodies with subatomic
  wave-function resolution,'' \emph{Physical Review B}, vol.~92, no.~8, p.
  085301, 2015.

\bibitem{heyd2003hybrid}
J.~Heyd, G.~E. Scuseria, and M.~Ernzerhof, ``Hybrid functionals based on a
  screened coulomb potential,'' \emph{The Journal of chemical physics}, vol.
  118, no.~18, pp. 8207--8215, 2003.

\bibitem{AhmedTFET}
\BIBentryALTinterwordspacing
S.~Z. Ahmed, Y.~Tan, D.~S. Truesdell, B.~H. Calhoun, and A.~W. Ghosh,
  ``Modeling tunnel field effect transistors—from interface chemistry to
  non-idealities to circuit level performance,'' \emph{Journal of Applied
  Physics}, vol. 124, no.~15, p. 154503, 2018. [Online]. Available:
  \url{https://doi.org/10.1063/1.5044434}
\BIBentrySTDinterwordspacing

\bibitem{InAlAs_MC}
\BIBentryALTinterwordspacing
J.~Zheng, Y.~Yuan, Y.~Tan, Y.~Peng, A.~Rockwell, S.~R. Bank, A.~W. Ghosh, and
  J.~C. Campbell, ``Simulations for inalas digital alloy avalanche
  photodiodes,'' \emph{Applied Physics Letters}, vol. 115, no.~17, p. 171106,
  2019. [Online]. Available: \url{https://doi.org/10.1063/1.5114918}
\BIBentrySTDinterwordspacing

\bibitem{dunn1997monte}
G.~Dunn, G.~Rees, J.~David, S.~Plimmer, and D.~Herbert, ``Monte carlo
  simulation of impact ionization and current multiplication in short gaas
  diodes,'' \emph{Semiconductor science and technology}, vol.~12, no.~1, p.
  111, 1997.

\bibitem{zheng2016pmt}
J.~Zheng, L.~Wang, X.~Wu, Z.~Hao, C.~Sun, B.~Xiong, Y.~Luo, Y.~Han, J.~Wang,
  H.~Li \emph{et~al.}, ``A pmt-like high gain avalanche photodiode based on
  gan/aln periodically stacked structure,'' \emph{Applied Physics Letters},
  vol. 109, no.~24, p. 241105, 2016.

\bibitem{GangulyPbSe}
\BIBentryALTinterwordspacing
S.~Ganguly, M.-H. Jang, Y.~Tan, S.-S. Yoo, M.~C. Gupta, and A.~W. Ghosh, ``A
  multiscale materials-to-systems modeling of polycrystalline pbse
  photodetectors,'' \emph{Journal of Applied Physics}, vol. 126, no.~14, p.
  143103, 2019. [Online]. Available: \url{https://doi.org/10.1063/1.5087818}
\BIBentrySTDinterwordspacing

\bibitem{huntington2020ingaas}
A.~S. Huntington, \emph{InGaAs Avalanche Photodiodes for Ranging and
  Lidar}.\hskip 1em plus 0.5em minus 0.4em\relax Woodhead Publishing, 2020.

\bibitem{yuan2018temperature}
Y.~Yuan, J.~Zheng, Y.~Tan, Y.~Peng, A.-K. Rockwell, S.~R. Bank, A.~Ghosh, and
  J.~C. Campbell, ``Temperature dependence of the ionization coefficients of
  inalas and algaas digital alloys,'' \emph{Photonics Research}, vol.~6, no.~8,
  pp. 794--799, 2018.

\end{thebibliography}

%




\end{document}